\documentclass[prd,preprint,nofootinbib]{revtex4-1} 

\usepackage{latexsym,verbatim,enumerate}
\usepackage{amssymb}
\usepackage{amsfonts}
\usepackage{amsmath,color}
\usepackage{graphicx}
\usepackage{bm}
\usepackage[utf8]{inputenc}
\usepackage[T1]{fontenc}

\usepackage{hyperref}

\hypersetup{
    colorlinks=true,
    linkcolor=red,
    citecolor=red,
    filecolor=red,      
    urlcolor=red,
    pdftitle={On the catastrophe time of fluids under the action of a gravitational field},
    pdfpagemode=FullScreen,}

\def\ber{\begin{eqnarray}}
\def\eer{\end{eqnarray}}
\def\beq{\begin{equation}}
\def\eeq{\end{equation}}

\def\ed{\end{document}}
\def\di#1#2{\frac{\mathrm{d} #1}{\mathrm{d}#2}}

\def\dt#1{\frac{\mathrm{d} #1}{\mathrm{d}t}}

\newcommand{\ppar}[2]{\frac{\partial #1}{\partial #2}}

\def\di#1#2{\frac{\mathrm{d} #1}{\mathrm{d}#2}}

\makeatletter
\let\jnl@style=\rm
\def\ref@jnl#1{{\jnl@style#1}}

\def\aj{\ref@jnl{AJ}}                   
\def\actaa{\ref@jnl{Acta Astron.}}      
\def\araa{\ref@jnl{ARA\&A}}             
\def\apj{\ref@jnl{ApJ}}                 
\def\apjl{\ref@jnl{ApJ}}                
\def\apjs{\ref@jnl{ApJS}}               
\def\ao{\ref@jnl{Appl.~Opt.}}           
\def\apss{\ref@jnl{Ap\&SS}}             
\def\aap{\ref@jnl{A\&A}}                
\def\aapr{\ref@jnl{A\&A~Rev.}}          
\def\aaps{\ref@jnl{A\&AS}}              
\def\azh{\ref@jnl{AZh}}                 
\def\baas{\ref@jnl{BAAS}}               
\def\bac{\ref@jnl{Bull. astr. Inst. Czechosl.}}
\def\caa{\ref@jnl{Chinese Astron. Astrophys.}}
\def\cjaa{\ref@jnl{Chinese J. Astron. Astrophys.}}
\def\icarus{\ref@jnl{Icarus}}           
\def\jcap{\ref@jnl{J. Cosmology Astropart. Phys.}}
\def\jrasc{\ref@jnl{JRASC}}             
\def\memras{\ref@jnl{MmRAS}}            
\def\mnras{\ref@jnl{MNRAS}}             
\def\na{\ref@jnl{New A}}                
\def\nar{\ref@jnl{New A Rev.}}          
\def\pra{\ref@jnl{Phys.~Rev.~A}}        
\def\prb{\ref@jnl{Phys.~Rev.~B}}        
\def\prc{\ref@jnl{Phys.~Rev.~C}}        
\def\prd{\ref@jnl{Phys.~Rev.~D}}        
\def\pre{\ref@jnl{Phys.~Rev.~E}}        
\def\prl{\ref@jnl{Phys.~Rev.~Lett.}}    
\def\pasa{\ref@jnl{PASA}}               
\def\pasp{\ref@jnl{PASP}}               
\def\pasj{\ref@jnl{PASJ}}               
\def\rmxaa{\ref@jnl{Rev. Mexicana Astron. Astrofis.}}%
\def\qjras{\ref@jnl{QJRAS}}             
\def\skytel{\ref@jnl{S\&T}}             
\def\solphys{\ref@jnl{Sol.~Phys.}}      
\def\sovast{\ref@jnl{Soviet~Ast.}}      
\def\ssr{\ref@jnl{Space~Sci.~Rev.}}     
\def\zap{\ref@jnl{ZAp}}                 
\def\nat{\ref@jnl{Nature}}              
\def\iaucirc{\ref@jnl{IAU~Circ.}}       
\def\aplett{\ref@jnl{Astrophys.~Lett.}} 
\def\apspr{\ref@jnl{Astrophys.~Space~Phys.~Res.}}
\def\bain{\ref@jnl{Bull.~Astron.~Inst.~Netherlands}}
\def\fcp{\ref@jnl{Fund.~Cosmic~Phys.}}  
\def\gca{\ref@jnl{Geochim.~Cosmochim.~Acta}}   
\def\grl{\ref@jnl{Geophys.~Res.~Lett.}} 
\def\jcp{\ref@jnl{J.~Chem.~Phys.}}      
\def\jgr{\ref@jnl{J.~Geophys.~Res.}}    
\def\jqsrt{\ref@jnl{J.~Quant.~Spec.~Radiat.~Transf.}}
\def\memsai{\ref@jnl{Mem.~Soc.~Astron.~Italiana}}
\def\nphysa{\ref@jnl{Nucl.~Phys.~A}}   
\def\physrep{\ref@jnl{Phys.~Rep.}}   
\def\physscr{\ref@jnl{Phys.~Scr}}   
\def\planss{\ref@jnl{Planet.~Space~Sci.}}   
\def\procspie{\ref@jnl{Proc.~SPIE}}   

\newtheorem{theorem}{Theorem}[section]
\newtheorem{remark}[theorem]{Remark}

\makeatother

\begin{document}

\author{Davide Astesiano}
\email{davide.astesiano@venturilab.ch}
\affiliation{Mathematics Division, Venturi Lab SA, Route du Pâqui 1, 1720 Corminboeuf, Switzerland}
\affiliation{Science Institute, University of Iceland,
Dunhaga 3, 107 , Reykjav\'{\i}k, Iceland}

\author{Giovanni Ortenzi}
\email{giovanni.ortenzi@unito.it
}
\affiliation{Dipartimento di Matematica ``G.Peano'', Universit\`a degli studi di Torino, Via Carlo Alberto 10, 10123 Torino, Italy}
\affiliation{INFN - Sezione di Torino , Via Pietro Giuria 1, 1025 Torino, Italy}

\author{Matteo Luca Ruggiero}
\email{matteoluca.ruggiero@unito.it}
\affiliation{Dipartimento di Matematica ``G.Peano'', Universit\`a degli studi di Torino, Via Carlo Alberto 10, 10123 Torino, Italy}
\affiliation{INFN - Sezione di Torino , Via Pietro Giuria 1, 1025 Torino, Italy}

\date{\today}

\title{On the catastrophe time of fluids under the action of a gravitational field}

\begin{abstract}
Motivated by the central role of the Zel'dovich approximation in the description of cosmic structure formation through gravitational collapse, we investigate Burgers-type dynamics in a spherically symmetric gravitational field. In the Newtonian setting, we derive perturbatively the catastrophe time for radial motion by imposing the loss of invertibility of the Lagrangian map. We show that the perturbative expansion is controlled by the dimensionless parameter
$
\alpha=\mu/{r_0^3 v_0(r_0)'^2},
$
rather than by the local gravitational acceleration alone. Hence, the expansion remain valid even when gravity is strong. We then extend the analysis to radial geodesic motion in Schwarzschild spacetime. 
\end{abstract}

\maketitle

\section{Introduction} \label{sec:intro}

One of the important issue of current study in theoretical and observational cosmology is the formation of large-scale structures \cite{peebles}, which are supposed to emerge from the amplification of gravitational instabilities of small primordial fluctuations. In this context, in a seminal paper \cite{zeldovich} Zel’dovich, suggested a simple model to explain the formation of these structures. Accordingly, he considered the post-baryon-photon decoupling Universe as a rarefied medium composed of collisionless, dust-like particles with negligible pressure, interacting solely through Newtonian gravity. Within this approximation, particles follow straight-line trajectories, leading to the formation of caustics where the mass density becomes very large. However, the pancake-like caustics predicted by this model do not exhibit the correct gravitational dynamics: they rapidly disperse and therefore do not persist for long. A further refinement led to the so-called adhesion model \cite{zeldovich2, gurbatov2012large}, in which particles that cross each other are assumed to stick together. The  adhesion model corresponds to a multi-dimensional Burgers equation in the inviscid limit \cite{villone}; we know that Burgers equation in the inviscid limit produces shocks along surfaces (in three dimensions) on which the density is infinite and across which the velocity is discontinuous. The adhesion model can be formulated as a multi-dimensional Burgers equation in the inviscid limit \cite{villone}.  
 The inviscid Burgers (Hopf) equation develops shock structures which, in multidimensions, form along hypersurfaces where the density formally diverges and the velocity field becomes discontinuous across them (see e.g. \cite{bressan2025generic} for a mathematical approach of the 2D case). The systematic study and classification of singularities developed in finite times is an active field of research  also from the point of view of differentiable mappings \cite{vi82singularities}.
 In fluid dynamics context, the collective motion of collisionless particles is modeled by pressureless fluids, called incoherent fluids or dust \cite{landau1987fluid}. 
The study of finite time singularity formation for incoherent fluids is an ubiquitous phenomenon which 
typically ranges from geophysical scales  to gravitational phenomena in astrophysics  \cite{chefranov1991exact, zubarev2018exact, kuznetsov2003towards, kuznetsov2022slipping, konopelchenko2025euler}. The characteristic times or  structures of the fluid singularities is an hint of real structures where the pressure is present.
 In General Relativity  a lot of work has been done in the study of light-like  caustics   (see e.g. \cite{wambsganss1998gravitational, straumann2012general} and references therein regarding the gravitational lensing) while their massive counterpart seems less studied.

In this paper, we investigate inviscid Burgers dynamics in a spherically symmetric gravitational field. Starting from a Newtonian framework, we employ a perturbative approach to compute the blow-up time for radially infalling particles. We then extend the analysis to a general relativistic setting by considering radial geodesic motion in Schwarzschild spacetime.

The paper is organized as follows: after revising the basic idea of Burgers dynamics for unidimensional models in Section \ref{sec:simple}, we study radial motion in Newtonian gravity in Section \ref{sec:radial} and then we extend our approach to a general relativistic framework in Section \ref{sec:sch}. Eventually, we discuss our results in Section \ref{sec:disc}.

\section{Simple unidimensional models}\label{sec:simple}

Here we briefly recall the relevant features of Burgers dynamics for free particles and uniformly accelerated ones, before considering the general case of a Newtonian potential in the following Section.

In one dimension, the inviscid Burgers equation can be written as
\beq
\partial_{t}v+v\partial_{x}v=0 \label{eq:burg1}
\eeq
which, introducing the material derivative $\displaystyle \dt{}=\partial_{t}+v\partial_{x}$, becomes
\beq
\dt{v}=0 \label{eq:burg2}
\eeq
Equation (\ref{eq:burg1}) represents the \textit{Eulerian} formulation of the dynamics, where $v=v(x,t)$ is the velocity field and the fluid motion is described locally at fixed spatial position $x$. In this description, particles are not individually tracked during the evolution.

On the other hand, using a \textit{Lagrangian} approach, the motion of the fluid is described by the flow map $X=X(p,t)$, which gives the location at time $t$ of the particle whose initial position is $p$. Accordingly, the dynamics is determined by
\beq
\dt{X}=v\left(X(p,t),t \right) \label{eq:burg3}
\eeq
where, in this case, particles are identified by their initial position and the motion is described as a deformation of the initial domain.

The two descriptions are equivalent as long as the flow map is invertible, i.e. it is a diffeomorphism. In fact, from (\ref{eq:burg3}) we can set
\beq
v(x,t)=\dt{X}(p,t), \ \mathrm{with\ \ } p=X^{-1}\left(x,t\right) \label{eq:burg4}
\eeq

To illustrate when this equivalence breaks down, consider the Eulerian problem  (\ref{eq:burg1})–(\ref{eq:burg2}) with initial condition $v(x,0)=v_{0}(x)$. The particle trajectories coincide with the characteristic curves of the Burgers equation, which satisfy
\beq
\dt X=v(X,t), \quad X(p,0)=p \label{eq:burg5}
\eeq
Since the material derivative of the velocity vanishes, the velocity is constant along particle trajectories. Hence  $v\left(X(p,t),t\right)=v_{0}(p)$, from (\ref{eq:burg3}) we obtain
\beq
\dt{X}=v_{0}(p) \label{eq:burg6}
\eeq
which can be immediately integrated to give
\beq
X(p,t)=p+v_{0}(p)t \label{eq:burg7}
\eeq

The flow map remains invertible as long as its Jacobian $J(p,t)=\partial_{p}X(p,t)$ does not vanish. Since $J(p,t)=1+\partial_{p}v_{0}(p)t$,   the shock formation corresponds to the \textit{catastrophe  or blow-up  time} $\bar t$ for which $J(p,\bar t)=0$, which reads
\beq
1+\partial_{p}v_{0}(p)\bar t=0 \label{eq:burg8}
\eeq
or, more precisely,
\beq
\bar t=\inf_{p:\partial_{p}v_{0}(p)<0}\left(-\frac{1}{\partial_{p}v_{0}(p)}\right) \label{eq:burg88}
\eeq
At this time the flow map ceases to be invertible and a singularity forms, corresponding to the intersection of particle trajectories, and in addition, the velocity gradient diverges, $\partial_x v\to -\infty$, signalling the formation of a shock.

As a simple example, consider $v_{0}(p)=-p$. In this case we simply obtain $X(p,t)=p\left(1-t\right)$; the characteristic curves are straight lines in the $(x,t)$ plane: $x=p\left(1-t \right)$, and each value of $p$ defines a specific evolution of the particle motion. We see that as long as $t< \bar t=1$, the particles do not intersect, but they converge to $x=0$ when $t=1$. This toy model elucidates the key features of the phenomenon under consideration.

A more realistic situation for our purposes is described by the following generalization of Eq. (\ref{eq:burg1})
\beq
\partial_{t}v+v\partial_{x}v=g \label{eq:burg11}
\eeq
which corresponds to the motion of the fluid particles in a \textit{constant} field $g$, which approximates the gravitational field in a sufficiently small region of space around the Earth. Using the same approach described above, we obtain the following results for the flow map
\beq
X(p,t)=p+v_{0}(p)t+\frac 1 2 g t^{2}, \quad v\left(X(a,t),t \right)=v_{0}(a)+gt \label{eq:burg12}
\eeq

Since $g$ does not depend on $p$, the condition for the breakdown of invertibility of the flow map is the same as before, and the condition for blow-up is again given by Eq. (\ref{eq:burg8}). In addition, it is easy to check that Eq. (\ref{eq:burg11}) is mapped into the free evolution (\ref{eq:burg1}) by using the transformation
\beq
X=x-\frac 1 2 g t^{2}, \quad V(x,t)=v(x,t)-gt \label{eq:burg13}
\eeq
from which we obtain
\beq
\partial_{t}V+V\partial_{X}V=0 \label{eq:burg14}
\eeq
The transformation (\ref{eq:burg13}) describes the passage to a uniformly accelerated reference frame.

{
Still working  in one spatial dimension, the previous considerations can be easily generalized to the case of an incoherent fluid subject to an external field 
\begin{equation}
\begin{split}
&v_t+v v_x + U'(x)=0\, , \\
&v(x,0)=v_0(x)\, .
\end{split}
\label{eq:BHsorce}
\end{equation}
where $U(x)$  defines the potential of an assigned source. The characteristic curves of the previous system are defined by the Hamilton equations
\begin{equation}
\begin{split}
&\frac{\mathrm{d} x}{\mathrm{d} t} = v\, , \qquad \frac{\mathrm{d} v}{\mathrm{d} t} =-\frac{\partial U(x)}{\partial x}\, , \\
&x(0)=x_0  \, , \qquad v(x(0),0)= v_0(x_0)\, .
\end{split}
\end{equation}
Additionally, the system admits the conserved quantity
\begin{equation}
E=\frac{1}{2}v^2+U(x)=\frac{1}{2}v_0(x(0))^2+U(x(0))=E(x_0)
\label{eq:genene}
\end{equation}
so that the characteristic curves are readily obtained from
\begin{equation}
t=t(x; x_0, E(x_0))=\int_{x_0}^x \frac{\mathrm{d} s}{\sqrt{2(E-U(s))}}\, .
\label{eq:genchar}
\end{equation}
Two different characteristics starting from $x_0$ and $x_0+\delta$ intersect each other in the spacetime generating a gradient blowup, if for the same value of $t$ and $x$ it holds
\begin{equation}
t(x; x_0, E(x_0))=t(x; x_0+\delta, E(x_0+\delta))\, .
\end{equation}
In the limit of small $\delta$ the previous condition becomes
\begin{equation}
\di{t}{x_{0}}=\frac{\partial t}{\partial x_0}+\ppar{t}{E}\frac{\partial E}{\partial x_0}\equiv 0\, .
\label{eq:genbucurve}
\end{equation}
The intersection of the relations (\ref{eq:genchar}) and (\ref{eq:genbucurve}) generates a curve in the spacetime $t=t(x)$, called blowup curve, which identifies for every characteristic the first blowup (see figure \ref{fig-catcurgen} for a qualitative sketch). 
\begin{figure}[!ht]
\begin{center}
\includegraphics[width=.5 \textwidth]{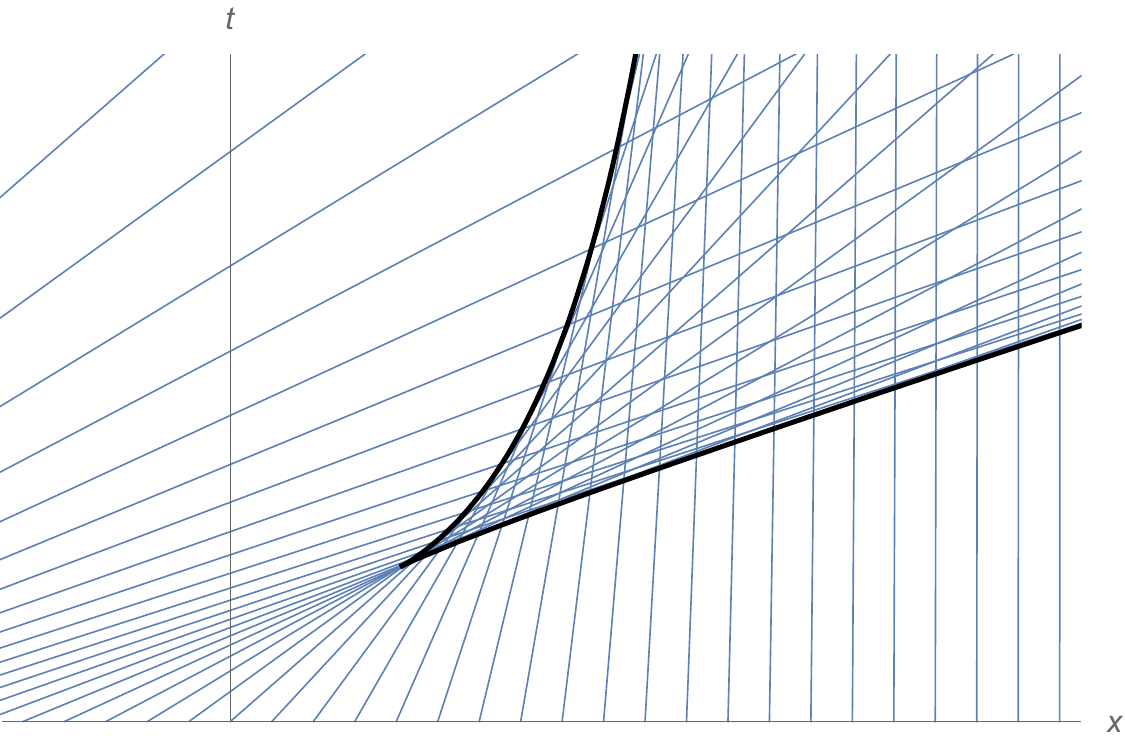}
\caption{Qualitative plot of  characteristics (\ref{eq:genchar}) (thin lines) and the blowup curve (\ref{eq:genbucurve}) (thick line). 
The intersection between a  characteristic and  the blowup curve gives the first point in the spacetime when the characteristics intersect a different characteristics.}
\label{fig-catcurgen}
\end{center}
\end{figure}

In the approximation (\ref{eq:burg12}) of a constant gravitational field, for example, 
the blowup curve generated by (\ref{eq:genchar}) and (\ref{eq:genbucurve})  is 

 \begin{equation}
 (x,t)= \left( x_0 -\frac{g}{2 (v_0'(x_0))^2} -\frac{v_0(x_0)}{v_0'(x_0)}, -\frac{1}{v_0'(x_0)}\right)
 \label{eq:buccg}
 \end{equation}
 and its shape depends on the acceleration $g$.
In figure \ref{fig-constacccone} is depicted an example of such dependence in the case of the initial value $v_0(x_0)=\exp(-x_0^2)$.
\begin{figure}[!ht]
\begin{center}
\includegraphics[width=.5 \textwidth]{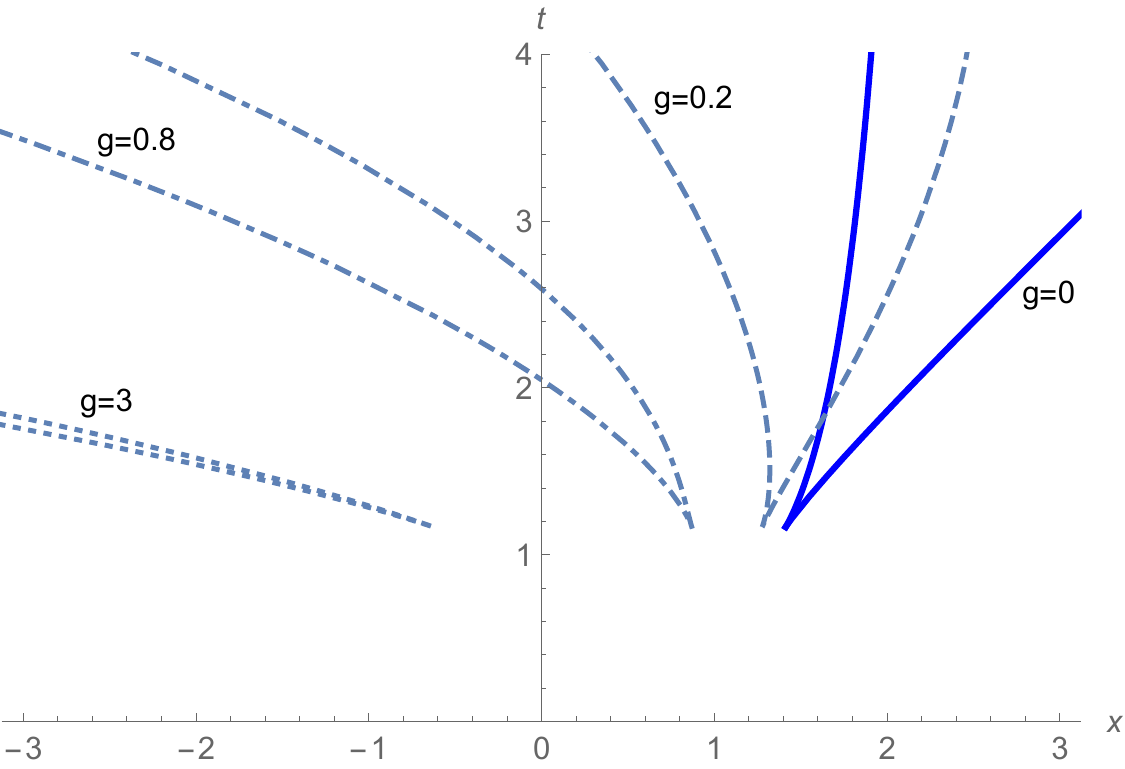}
\caption{The blowup curve (\ref{eq:buccg}) for the toy-model case of a constant gravitational field $g$ in the case of initial velocity value $v_0(x_0)=\exp(-x_0^2)$.}
\label{fig-constacccone}
\end{center}
\end{figure} 

In the free case of the inviscid Burgers equation the solution of the system can be written in hodograph form aa
\begin{equation}
x-ut=f(u)\, ,
\end{equation}
where $f$ is an arbitrary function. 
The analogue construction for the equation (\ref{eq:BHsorce}) is technically similar even if the result appears more involved.
Along every characteristics one has two conserved quantities: the energy $E(r_0)$ (see (\ref{eq:genene})) and the following quantity
derived from (\ref{eq:genchar}) 
\begin{equation}
\Psi(r_0)=t \pm \int \frac{\mathrm{d}s }{ \sqrt{\frac{1}{2} u^2 +U(r) - U(s)}}\, .
\end{equation}
where now we refer to a unidimensional motion along the $r$ coordinate, since in what follows we are interested in radial motion in a gravitational field.
Generally, it is not possible to isolate $r_0$ from one conserved quantity and substitute in the other: however for a generic function $f$ it holds
\begin{equation}
F(u,r,t)\equiv E(u,r)-f[\Psi(u,r,t)]= 0
\label{eq:genhodo}
\end{equation}
where $f$ is an arbitrary function,
\[
E(u,r)={\frac{1}{2} u^2 +U(r)}
\]
and
\[
\Psi(u,r,t)=t \pm \int^r \frac{\mathrm{d}s }{ \sqrt{2(\frac{1}{2} u^2 +U(r) - U(s)})}.
\]

To identify the gradient catastrophes in the hodograph context we must set
\begin{equation}
u_r = -\frac{F_r}{F_u} \to \infty\, ,
\end{equation}
which is generically equivalent to $F_u(u,r,t)=0$.  The first time when the gradient blows up is a critical point (minimium) of the relation $t=t(u,r)$ 
implicitely given by  the relation $F_u(u,r,t)=0$. By the inverse function theorem $t_u=-F_{uu}/F_{ut}$ and $t_r=-F_{ur}/F_{ut}$ we have 
\begin{equation}
F_{uu}=0\, , \qquad F_{ur}=0\, .
\end{equation}
At the catastrophe time $t=t_c$ we develop $u=u_c+\Delta u$\, $r=r_c+\Delta r$ and we obtain
\begin{equation}
\begin{split}
0=&F(t_c,u_c+\Delta u,r_c+\Delta r)=\\
=&F^c+  F_r^c \Delta r +F_u^c \Delta u +\frac{1}{2} F_{rr}^c (\Delta r)^2 + F_{ur}^c \Delta r \Delta u + 
\frac{1}{2}  F_{uu}^c (\Delta u)^2 + \frac{1}{6} F_{uuu}^c (\Delta u)^3 +\dots\, .
\end{split}
\end{equation}
where $f^c=f(t_c,u_c,r_c)$.
the conditions $F_u^c=F_{uu}^c=F_{ur}^c=0$ imply, at lowest order,
\begin{equation}
\Delta u \sim  \left( - \frac{6 F^c_{r}}{F^c_{uuu}} \Delta r \right)^{1/3}
\end{equation}

In the previous example of constant gravitational field $U(s)=g s$  the hodograph equation (\ref{eq:genhodo}) becomes
\begin{equation}
 {\frac{1}{2} u^2 +g r} = f \left( u+gt\right)\, ,
 \label{eq:hodoconst}
\end{equation}
where $f$ is an arbitrary function.
The catastrophe conditions are 
\begin{equation}
\frac{1}{2} u_c^2 +g r_c - f \left( u_c+gt_c\right)=0\, , \qquad u_c-f'(u_c+gt_c)=0\, , \qquad 1-f''(u_c+gt_c)=0\, ,
\end{equation}
and the behavior near to the catastrophe  at $t=t_c$ is
\begin{equation}
\Delta u \sim \left( \frac{6g}{f'''(u_c+gt_c)} \Delta r\right)^{1/3}\, .
\end{equation}

This singular behavior is the generalization, in presence of a constant gravitational field $g$, of the classical 
Gurevich-Pitaevskij universal behavior for the free $g=0$ case  \cite{gurevich1974nonstationary, konopelchenko2024pressureless}. 

\begin{remark}
The limit $g \to 0$ appears degenerate because the hodograph form (\ref{eq:hodoconst}) is singular at $g =0$. 
It is more obvious if one use the equivalent form
\begin{equation}
 {\frac{1}{2} u^2 +g r} = f \left( r-ut -\frac{1}{2}gt^2\right)\, .
\end{equation}
\end{remark}}
The explicit computations carried out in the case of a constant gravitational field become significantly more involved in the general setting. In the following Sections, we calculate the catastrophe time for particles moving in more realistic gravitational fields. In particular, we first consider the Newtonian gravitational field and subsequently its general relativistic counterpart described by the Schwarzschild solution

\section{Radial Newtonian Motion}\label{sec:radial}
We focus on  the  Newtonian field and consider one-dimensional radial motion, which provides a natural generalization of Eq. (\ref{eq:burg11}) to a non-uniform gravitational field. In terms of the coordinates $(r,t)$, the model equation is
\beq
\partial_{t} v+v\partial_{r}v=-\frac{\mu}{r^{2}} \label{eq:new_rad1}
\eeq
where $v=v(r,t)$ is the radial velocity, and we set $\mu=GM$, with $M$ denoting the mass of the source of the gravitational field.

As before, the particle trajectories coincide with the characteristic curves of the Burgers equation. Setting $v(r,t)=\dot r$, these trajectories can be obtained from the conservation of energy. In particular, since the motion is radial and the angular momentum vanishes,
\beq
\dot r^{2}=2E+2\frac{\mu}{r}\, . \label{eq:en2}
\eeq

To compute the catastrophe time, we use a perturbative approach. We assume that the first particle starts from the position $r_0$ with velocity $v_0$, while the second particle starts from $r_0+\epsilon$ with velocity $v_0+\epsilon v_0^{'}$
\footnote{Accordingly, $v_{0}'$ has the dimension of an inverse time.}. We then impose that the two particles reach the same position $\bar r$ at the same time $\bar t$. 

Let us first perform a simple calculation to infer the structure of the expansion. Roughly speaking, if the particle is sufficiently far from the source, the equation of motion can be estimated as
\begin{align}
    r_{r_0,v_0}(t)= r_0 + v_0 t-\frac{\mu}{2 r_0^2} t^2+O(r_0^{-1},v_0^{2}).
\end{align}
The second particle can similarly be described by
\begin{align}
    r_{r_0+ \epsilon ,v_0 +\epsilon v_0^{'}}(t)= r_0+ \epsilon + (v_0 +\epsilon v_0^{'}) t-\frac{\mu}{2 (r_0+ \epsilon)^2} t^2+O(r_0^{-1},v_0^{2},\epsilon^2).
\end{align}
The particles reach the same point if
\begin{align}
 \bar r_{r_0,v_0}(\bar t)-\bar r_{r_0+\epsilon,v_0 +\epsilon v_0^{'} }(\bar t)=0.   
\end{align}
We remark that this corresponds to the Lie derivative along the flow, which, for a scalar function, reduces to an ordinary derivative. At first order in $\epsilon$, the solution is
\begin{align}
    \bar t=-\frac{1}{v_0^{'}}-\frac{\mu }{r_0^3 v_0^{' 3}}-\frac{2 \mu ^2}{r_0^6 v_0^{' 5}}.
\end{align}
As we shall see, this simple calculation correctly identifies the expansion parameter, although it does not give the correct coefficients. Motivated by the estimate above, we define
\begin{align} \label{eq:param}
    \alpha:=\frac{\mu}{r_0^3 v_0^{'2}},  \qquad \beta:=\frac{v_0}{r_0 v_0^{'}}.
\end{align}

We now treat the problem by expanding in the parameter $\alpha$, and we extract the corresponding series systematically. By introducing the parameters above, Eq. (\ref{eq:en2}) can be rewritten as
\begin{align}
   \left( \frac{dR(T)}{dT}\right)^2&=1-2 \frac{\alpha}{\beta^2} [1-R(T)^{-1}], \nonumber \\
    R&= r/r_0, \qquad T= \beta v_0^{'} t.  
\end{align}
This expression clearly admits an expansion in powers of the small parameter $\alpha$. The solution can therefore be written perturbatively as
\begin{align}
   R=1 + T + \frac{\alpha}{\beta^{2}}\left[\log(1+T) - T\right]+ O(\alpha^2).
   \label{eq:R_t_pert_Newton}
\end{align}
In terms of the physical variables, this becomes
\begin{equation}
r(t)
=
r_0+v_0t
+\mu\left[
-\frac{t}{r_0 v_0}
+\frac{1}{v_0^2}\log\!\left(\frac{r_0+v_0t}{r_0}\right)
\right]
+O(\mu^2).
\label{eq:r_t_pert_Newton}
\end{equation}

To determine the first catastrophe time, we impose the condition
\begin{align}
    \frac{d}{dr_0} r(t,r_0,v_0(r_0))\vert_{t=\bar t}=0.
\end{align}
The solution, written in terms of $\alpha$, is
\begin{align}
   \bar t=-\frac{1}{v_0^{'}}+\frac{\alpha}{v_0^{'}} \frac{  \beta  (\beta +2)+2 \log (1-\beta )}{\beta ^3 }+O( \alpha^2),
\end{align}
while in terms of the physical variables it reads
\begin{align}
    \bar t=-\frac{1}{v_0^{'}}+\frac{\mu}{v_0^3}  \left[2 \log \left(1-\frac{v_0}{r_0 v_0^{'}}\right)+\frac{v_0 (2 r_0 v_0^{'}+v_0)}{r_0^2 v_0^{'2}}\right]+O( \alpha^2). \label{blowup1}
\end{align}
To compare this expression with the preliminary estimate above, we expand it in inverse powers of $r_0$, obtaining
\begin{align}
    \bar t=-\frac{1}{v_0^{'}}-\frac{2 \mu }{3 r_0^3v_0^{'3}}-\frac{\mu  v_0}{2 r_0^4 v_0^{'4}}-\frac{2 \mu  v_0^{2}}{5 r_0^5 v_0^{'5}}+ O(r^{-8}, v^{3}).
\end{align}

This formalism can be extended straightforwardly to higher orders in $\alpha$. In fact, the next terms are
\begin{align}
   \bar t&=-\frac{1}{v_0^{'}}+\frac{\alpha}{v_0^{'}} \frac{  \beta  (\beta +2)+2 \log (1-\beta )}{\beta ^3 }\nonumber\\
   &+\frac{\alpha ^2 \left(-\beta ^4-\beta ^3-12 \beta +6 \beta  \log (1-\beta )-12 \log (1-\beta
   )\right)}{(\beta -1) \beta ^5  v_0^{'}} +\nonumber\\
   &+\frac{\alpha ^3 \left(2 \beta ^6-4 \beta ^4-12 \beta ^3-93 \beta ^2+18 \beta ^2 \log (1-\beta )\right)}{2 (\beta -1)^2 \beta ^7
   v_0^{'}}+\nonumber\\
   &+\frac{\alpha ^3 \left(90 \beta -18
   \log ^2(1-\beta )-156 \beta  \log (1-\beta )+90 \log (1-\beta )\right)}{2 (\beta -1)^2 \beta ^7
   v_0^{'}}+\nonumber\\
   &+O( \alpha^5),
\end{align}
where we have assumed $v_0^{'}<0$.

We emphasize that the perturbative expansion is not an expansion in the local gravitational acceleration $\mu/r_0^2$ itself. Rather, the relevant dimensionless parameter is $\alpha$.
Therefore, the expansion remains meaningful even when the gravitational acceleration at the initial position is large, provided that $\alpha$ is sufficiently small. In this sense, the approximation is controlled by the ratio between the gravitational scale and the scale set by the initial velocity gradient. Thus, by varying $\alpha$, the perturbative solution interpolates between weak-gravity regimes and regimes in which the local gravitational field can be large, while still keeping the expansion under control.

\section{Radial Motion In Schwarzschild Spacetime}\label{sec:sch}
The spacetime interval for the Schwarzschild geometry \cite{rindler2012essential} can be written, for radial motion, in the form
\beq
ds^{2}=-A(r)c^2dt^{2}+\frac{dr^{2}}{A(r)} \label{eq:schmetric}
\eeq
where $A(r)=1-\frac{2\mu}{c^{2}r}$. We can write the Lagrangian as 
$L=\frac 1 2 g_{\mu\nu}\dot{x}^{\mu}\dot{x}^{\nu}$; in this section, a dot denotes differentiation with respect to the proper time $c\, d\tau=\sqrt{-ds^{2}}$ of the falling particle. The invariance of $L$ under coordinate-time translations gives the constant of motion $E$, such that $\dot t=\frac{E}{A(r)}$. Accordingly, for massive particles ($L=-1/2$), we obtain
\beq
\frac{\dot r ^{2}}{c^2}=E^{2}-1+\frac{2\mu}{c^2 r }\label{eq:schrdot0}
\eeq
or, equivalently,
\beq
\dot r ^{2}=2\mathcal E+\frac{2\mu}{r} \label{eq:schrdot1}
\eeq
where we have set $2\mathcal E/c^2=E^{2}-1$; note that for ``closed'' orbits one has $E^{2}-1<0$. 

We see that Eq. (\ref{eq:schrdot1}) has the same form as the Newtonian Eq. (\ref{eq:en2}). Accordingly, in this context, we can interpret the solutions (\ref{eq:R_t_pert_Newton}) and 
(\ref{eq:r_t_pert_Newton}) as the parametric form of the radial coordinate and proper time in the metric (\ref{eq:schmetric}) for a test particle starting at $r=r_{0}$ with initial proper-time radial velocity $\frac{dr}{d\tau}(\tau=0):=u_{0}$. Therefore, the trajectory can be written, up to first order in $\mu$, as
\begin{equation}
r(\tau)
=
r_0+u_0\tau
+\mu\left[
-\frac{\tau}{r_0u_0}
+\frac{1}{u_0^2}\log\!\left(\frac{r_0+u_0\tau}{r_0}\right)
\right]
+O(\mu^2).
\label{eq:r_tau_pert}
\end{equation}
Instead of the parameters defined in Eq. (\ref{eq:param}), we now introduce their relativistic counterparts,
\begin{align} \label{eq:paramrel}
    \alpha_r:=\frac{\mu}{r_0^3 u_0^{'2}},  \qquad \beta_r:=\frac{u_0}{r_0 u_0^{'}},
\end{align}
and, analogously,
\begin{align}
    R&= r/r_0, \qquad T_r= \beta_r u_0^{'} \tau=\frac{u_0}{r_0}\tau.  
\end{align}
The trajectory then takes the same form as before,
\begin{align}
   R=1 + T_{r} + \frac{\alpha_{r}}{\beta_{r}^{2}}\left[\log(1+T_{r}) - T_{r}\right]+ O(\alpha_r^2).
\end{align}

To compare the relativistic dynamics with the Newtonian result obtained in the previous Section, it is convenient to express the motion in terms of the Schwarzschild coordinate time $t$, rather than in terms of the proper time $\tau$. Starting from the conservation law
\begin{equation}
\dot t=\frac{E}{A(r)}, \qquad A(r)=1-\frac{2\mu}{c^2 r},
\label{eq:schw_tdot}
\end{equation}
and using the normalization condition for timelike geodesics, we can write
\beq
\frac{dt(\tau)}{d\tau}=\frac{E}{A(\tau)}=\frac{\sqrt{1-\frac{2\mu}{r_0 c^2}+\frac{u_0^2}{c^2} }}{1-\frac{2\mu}{r(\tau) c^2}} .
\label{eq:dtdtau1}
\eeq
As in the Newtonian case, we work perturbatively to first order in $\alpha_r$. Substituting \eqref{eq:r_tau_pert} into \eqref{eq:dtdtau1} and expanding to first order in $\alpha_r$, we obtain
\begin{equation}
\frac{dt}{d\tau}
=
\sqrt{1+\frac{u_0(r_0)^2}{c^2}}
\left[
1+\frac{\alpha_r}{\beta_r^2}\dfrac{u_0(r_0)^2}{c^2}\,
\frac{
\left(
1-T_r
+2\,\dfrac{u_0(r_0)^2}{c^2}
\right)}
{\left(1+T_r\right)
\left(1+\dfrac{u_0(r_0)^2}{c^2}\right)}
\right]
+O(\alpha_r^2).
\label{eq:dtdtau_expanded}
\end{equation}
Integrating and fixing the integration constant by imposing $t(0)=0$, we get
\begin{equation}
t(\tau)=
\sqrt{1+\frac{u_0(r_0)^2}{c^2}}\,
\tau\left[
1+\frac{\alpha_r}{\beta_r^2}\,
\dfrac{u_0(r_0)^2}{c^2}
\left(
2 \frac{\log\!\left(1+T_r\right)}{T_r}
-\frac{1}{1+\dfrac{u_0(r_0)^2}{c^2}}
\right)
\right]+O(\alpha_r^2).
\label{eq:t_of_tau}
\end{equation}
We now invert Eq.~\eqref{eq:t_of_tau} perturbatively. At first order in $\alpha_r$, we find
\begin{equation}
\tau(t)=
\frac{t}{\sqrt{1+\dfrac{u_0(r_0)^2}{c^2}}}
\left[
1+\frac{\alpha_r}{\beta_r^2}\dfrac{u_0(r_0)^2}{c^2}
\left(
\frac{1}{1+\dfrac{u_0(r_0)^2}{c^2}}
-
\frac{2 \sqrt{1+\dfrac{u_0(r_0)^2}{c^2}}}{\dfrac{u_0(r_0)}{r_0}\,
t} 
\log\!\left(
1+\frac{\dfrac{u_0(r_0)}{r_0}\,t}
{\sqrt{1+\dfrac{u_0(r_0)^2}{c^2}}}
\right)
\right)
\right]
\label{eq:tau_of_t}
\end{equation}
which indeed satisfies $t(\tau(t))=t+O(\alpha_r^2)$.
By composing \eqref{eq:r_tau_pert} with \eqref{eq:tau_of_t}, we obtain the trajectory in Schwarzschild coordinate time,
\begin{align}
r(t)=
r_0\left(1+\frac{\dfrac{u_0(r_0)}{r_0}\,t}{\sqrt{1+\dfrac{u_0(r_0)^2}{c^2}}}\right)
+\frac{\alpha_r}{\beta_r^2}\,r_0
\left[
\left(1-2\frac{u_0(r_0)^2}{c^2}\right)
\log\!\left(
1+\frac{\dfrac{u_0(r_0)}{r_0}\,t}{\sqrt{1+\dfrac{u_0(r_0)^2}{c^2}}}
\right)
-
\frac{\dfrac{u_0(r_0)}{r_0}\,t}
{\left(1+\dfrac{u_0(r_0)^2}{c^2}\right)^{3/2}}
\right].
\end{align}

At this stage, the expression is still written in terms of the initial proper-time velocity $u_0$. To compare with the Newtonian result, it is convenient to rewrite everything in terms of the initial coordinate three-velocity $v_{0}$ by using
\begin{equation}
u_0=
\frac{\sqrt{1-\dfrac{2\mu}{c^2r_0}}\,v_0}
{\sqrt{\left(1-\dfrac{2\mu}{c^2r_0}\right)^2-\dfrac{v_0^2}{c^2}}}.
\label{eq:u0_v0_relation}
\end{equation}
We can then write the coordinate-time trajectory in terms of the initial data $(r_0,v_0)$ as
\begin{align}
 r(t,v_0,r_0)=
r_0\left[
1+T
+\frac{\alpha}{\beta^2}
\left(1-3\frac{v_0(r_0)^2}{c^2}\right)
\bigl(\log(1+T)-T\bigr)
\right]+O(\alpha^2).
\end{align}
Here we have again used the non-relativistic parameters. Since the two perturbative parameters are related by
\begin{align}
    \alpha_r = \alpha \left(1-\frac{v_0^2}{c^2}\right)^3 + O(\alpha^2),\qquad
    \alpha = \alpha_r \left(1+\frac{u_0^2}{c^2}\right)^3 + O(\alpha_r^2),
\end{align}
the perturbative ordering is unchanged when passing from $\alpha_r$ to $\alpha$. In particular, an expansion at first order in $\alpha$ is also first order in $\alpha_r$, and vice versa. Therefore, rewriting the result in terms of the parameter $\alpha$ preserves the consistency of the relativistic expansion controlled by $\alpha_r$, independently of the value of the velocity.

We can now study the breakdown of invertibility of the Lagrangian map. As in the classical case, the catastrophe occurs when neighboring trajectories intersect, namely when the Jacobian
\begin{equation}
J(r_0,t)=\frac{\partial r(t,v_0(r_0),r_0)}{\partial r_0}
\end{equation}
vanishes.
We look for the first encounter time in the perturbative form
\begin{equation}
\bar t=\bar t^{(0)}+\alpha\,\bar t^{(1)}+O(\alpha^2).
\end{equation}
At zeroth order, the condition $J=0$ gives the familiar result
\begin{equation}
\bar t^{(0)}=-\frac{1}{v_0'}.
\end{equation}
After straightforward algebra, we obtain
\begin{align}
\bar t=
-\frac{1}{v_0'}
\left[
1-
\frac{\alpha}{\beta}
\left(
1+\frac{2}{\beta}
+\frac{2\log(1-\beta)}{\beta^2}
-
3\,\frac{v_0(r_0)^2}{c^2}
\right)
\right].
\end{align}
Equivalently, in explicit variables,
\begin{equation}
\bar t=
-\frac{1}{v_0'}
+\frac{\mu}{v_0^3}
\left[
2\log\!\left(1-\frac{v_0}{r_0v_0'}\right)
+\frac{v_0(2r_0v_0'+v_0)}{r_0^2 v_0'^2}
\right]
-\frac{3\mu v_0}{c^2 r_0^2 v_0'^2}
+O(\mu^2).
\label{RelTime}
\end{equation}
The additional term $\displaystyle \Delta t_{\mathrm{rel}}=-\frac{3\mu v_0}{c^2 r_0^2 v_0'^2}$ in Eq.~\eqref{RelTime} has a clear physical meaning. At leading order, the radial geodesic equation written in terms of the particle proper time has the same structure as the Newtonian energy equation. The difference appears when the crossing time is expressed in Schwarzschild coordinate time, namely in the time measured by a distant static observer. In this sense, the term $\Delta t_{\mathrm{rel}}$ encodes the combined effect of gravitational redshift and relativistic time dilation on the estimate of the catastrophe time. For infalling particles, $v_0<0$, one finds $\Delta t_{\mathrm{rel}}>0$, so that the first crossing is delayed with respect to the Newtonian prediction.

\section{Discussion}\label{sec:disc}
We have investigated the catastrophe time for radially free-falling particles in the presence of gravitational fields, both in the Newtonian and in the Schwarzschild setting. Starting from the corresponding equations of motion, we adopted a perturbative approach and derived the first crossing time by imposing the vanishing of the Lagrangian Jacobian. Although the explicit formulas obtained here have been computed only up to a finite order, the construction is systematic and can be extended, in principle, to higher perturbative orders.

A central point of the analysis is that the perturbative expansion is not controlled by the gravitational strength alone. The relevant dimensionless parameter is instead
\begin{equation}
\alpha=\frac{\mu}{r_0^3 v_0'^2},
\end{equation}
or its relativistic counterpart $\alpha_r$ defined by Eq. (\ref{eq:paramrel}), when the motion is parametrized by the proper time.

This means that the expansion may remain under control even when $\mu=GM$ itself, or the local gravitational acceleration $\mu/r_0^2$, is large. What matters is the comparison between the gravitational scale and the scale set by the initial velocity gradient. {More precisely, since $\displaystyle \frac{\mu}{r_{0}^{3}} \simeq \frac{1}{T^{2}_{N}}$, where $T_{N}$ is the scale of time of Newtonian free fall in the gravitational field, and $\displaystyle \frac{1}{v_{0}'}=T_{0}$ is the scale of time of free motion in absence of gravitational field, we see that 
\beq
\alpha \simeq \frac{T_{0}^{2}}{T_{N}^{2}} \label{eq:alphatimes}
\eeq
Accordingly, $\alpha$ is a measure of the relative importance of gravity in the formation of the caustic. Small values of $\alpha$ identify the perturbative regime: a relevant result that we found is the fact that the smallness of this parameter does not depend on gravity only, but on the comparison of the gravitational perturbation with the inertial motion, through their time scales.}

On the other hand,  increasing $\alpha$ requires the inclusion of higher-order corrections, such as the $O(\alpha^2)$ terms discussed above.

Within this framework, the Newtonian result provides the leading structure of the catastrophe time. In the Schwarzschild case, when the crossing time is expressed in terms of the coordinate time measured by a distant static observer, the final expression can be written as
\begin{equation}
\bar t = \bar t_{\mathrm{Newt}} + \Delta t_{\mathrm{rel}},
\end{equation}
where
\begin{equation}
\bar t_{\mathrm{Newt}}
=
-\frac{1}{v_0'}
+
\frac{\mu}{v_0^3}
\left[
2\log\!\left(1-\frac{v_0}{r_0 v_0'}\right)
+
\frac{v_0(2r_0 v_0' + v_0)}{r_0^2 v_0'^2}
\right]
\end{equation}
is the Newtonian contribution, while the leading Schwarzschild correction is
\begin{equation}
\Delta t_{\mathrm{rel}}
=
-\frac{3\mu v_0}{c^2 r_0^2 v_0'^2}.
\end{equation}
The condition $v_0'<0$ continues to select the regime in which neighboring trajectories approach each other and the Lagrangian map loses invertibility in finite time. The relativistic correction does not change this mechanism; rather, it modifies the time at which the caustic is observed in Schwarzschild coordinate time.

For infalling particles, $v_0<0$, the relativistic contribution is positive,
\begin{equation}
\Delta t_{\mathrm{rel}}>0,
\end{equation}
and therefore delays the first crossing with respect to the Newtonian prediction. This delay should not be interpreted as a weakening of the gravitational attraction. Instead, it reflects the fact that the catastrophe time is being measured in the coordinate time of a distant observer. The correction therefore encodes the combined effect of gravitational redshift and relativistic time dilation on the approach to the caustic.

Future work will aim to extend these results to more general free-falling motions, thereby relaxing the one-dimensional constraint.  The ongoing studies are devoted to singularities of stationary solutions. 
The presence of an external potential allows for nontrivial stationary solutions even in the  1+1D case (\ref{eq:new_rad1}), namely
\begin{equation}
v(r)=\pm \sqrt{k+\frac{\mu}{r}}\, , \qquad k \in \mathbb{R}.
\end{equation}
Away from the attraction center $r=0$, these solutions are regular. In particular, if $k<0$, the fluid is confined to the region $r<-\frac{k}{\mu}$. The extension  of such analysis in more dimensions suggests the existence of spacetime regions where the matter are confined. 

{ More generally, the study of this problem in two spatial dimensions allows one to consider motions with nonzero angular momentum \cite{prep}. In this setting, we expect the catastrophe condition to be governed not only by the local properties of the initial velocity field, but also by the spatial gradients of the relevant orbital invariants, namely the energy and the angular momentum. In particular, caustic formation should be associated with  the initial velocity profile, or equivalently with nontrivial variations in the initial energy and angular-momentum distributions. This suggests that the onset of singular behavior may be characterized in terms of the geometry of these invariant profiles and their induced mapping from the initial data to the evolved configuration.
}

Finally the methods developed in the paper  suggest the possibility to extend the results to  the study of light-like geodesics in the context, for example, of lensing problems. In particular the classical Ernst axially symmetric metrics \cite{PhysRev.167.1175} or the more recent cases studied  in  \cite{nbl9-6hnc,Astesiano:2022gph} seems a possible candidate to implement these techniques.  

\subsubsection*{Acknowledgments}
This project has received funding from  the PRIN 2022TEB52W-PE1 Project ``The charm of integrability: from nonlinear waves to random matrices". We also gratefully acknowledge the auspices of the GNFM Section of INdAM, under which part of this work was carried out, and the financial support of the project MMNLP (Mathematical Methods in Non Linear Physics) of the INFN.

\bibliographystyle{ieeetr} 
\bibliography{refs} 

\begin{thebibliography}{10}

\bibitem{peebles}
P.~J.~E. {Peebles}, {\em {The large-scale structure of the universe}}.
\newblock Princeton University Press, 1980.

\bibitem{zeldovich}
Y.~B. {Zel'dovich}, ``{Gravitational instability: An approximate theory for
  large density perturbations.},'' {\em \aap}, vol.~5, pp.~84--89, Mar. 1970.

\bibitem{zeldovich2}
S.~F. {Shandarin} and Y.~B. {Zeldovich}, ``{The large-scale structure of the
  universe: Turbulence, intermittency, structures in a self-gravitating
  medium},'' {\em Reviews of Modern Physics}, vol.~61, pp.~185--220, Apr. 1989.

\bibitem{gurbatov2012large}
S.~N. Gurbatov, A.~I. Saichev, and S.~F. Shandarin, ``Large-scale structure of
  the universe. the zeldovich approximation and the adhesion model,'' {\em
  Physics-Uspekhi}, vol.~55, no.~3, pp.~223--249, 2012.

\bibitem{villone}
U.~{Frisch}, J.~{Bec}, and B.~{Villone}, ``{Singularities and the distribution
  of density in the Burgers/adhesion model},'' {\em Physica D Nonlinear
  Phenomena}, vol.~152, pp.~620--635, May 2001.

\bibitem{bressan2025generic}
A.~Bressan, G.~Chen, and S.~Huang, ``Generic singularities for 2d pressureless
  flows,'' {\em Science China Mathematics}, vol.~68, no.~3, pp.~559--576, 2025.

\bibitem{vi82singularities}
A.~VI, S.~Guse{\"\i}n-Zade, and A.~Varchenko, ``Singularities of differentiable
  maps, vol. i (the classification of critical points, caustics and wave
  fronts),'' {\em Monographs in Mathematics}, vol.~82, 1982.

\bibitem{landau1987fluid}
L.~D. Landau and E.~M. Lifshitz, {\em Fluid Mechanics: Volume 6}, vol.~6.
\newblock Elsevier, 1987.

\bibitem{chefranov1991exact}
S.~Chefranov, ``An exact statistical closed description of vortex turbulence
  and of the diffusion of an impurity in a compressible medium,'' {\em Dokl.
  Akad. Nauk SSSR}, vol.~317, pp.~1080--1085, 1991.

\bibitem{zubarev2018exact}
N.~M. Zubarev and E.~Karabut, ``Exact local solutions for the formation of
  singularities on the free surface of an ideal fluid,'' {\em JETP Letters},
  vol.~107, no.~7, pp.~412--417, 2018.

\bibitem{kuznetsov2003towards}
E.~Kuznetsov, ``Towards a sufficient criterion for collapse in 3d euler
  equations,'' {\em Physica D: Nonlinear Phenomena}, vol.~184, no.~1-4,
  pp.~266--275, 2003.

\bibitem{kuznetsov2022slipping}
E.~Kuznetsov and E.~Mikhailov, ``Slipping flows and their breaking,'' {\em
  Annals of Physics}, vol.~447, p.~169088, 2022.

\bibitem{konopelchenko2025euler}
B.~G. Konopelchenko and G.~Ortenzi, ``On euler equation for incoherent fluid in
  curved spaces,'' {\em Physica D: Nonlinear Phenomena}, vol.~476, p.~134667,
  2025.

\bibitem{wambsganss1998gravitational}
J.~Wambsganss, ``Gravitational lensing in astronomy,'' {\em Living Reviews in
  Relativity}, vol.~1, no.~1, p.~12, 1998.

\bibitem{straumann2012general}
N.~Straumann, {\em General relativity}.
\newblock Springer Science \& Business Media, 2012.

\bibitem{gurevich1974nonstationary}
A.~Gurevich and L.~Pitayevsky, ``Nonstationary structure of a collisionless
  shock wave,'' {\em Soviet Journal of Experimental and Theoretical Physics},
  vol.~38, pp.~291--297, 1974.

\bibitem{konopelchenko2024pressureless}
B.~G. Konopelchenko and G.~Ortenzi, ``On pressureless euler equation with
  external force,'' {\em Physica D: Nonlinear Phenomena}, vol.~469, p.~134317,
  2024.

\bibitem{rindler2012essential}
W.~Rindler, {\em Essential relativity: special, general, and cosmological}.
\newblock Springer Science \& Business Media, 2012.

\bibitem{prep}
D.~Astesiano, G.~Ortenzi, and M.~L. Ruggiero, ``Jacobi fields and caustic
  formation in freely falling fluids,'' {\em in preparation}, 2026.

\bibitem{PhysRev.167.1175}
F.~J. Ernst, ``New formulation of the axially symmetric gravitational field
  problem,'' {\em Phys. Rev.}, vol.~167, pp.~1175--1178, Mar 1968.

\bibitem{nbl9-6hnc}
M.~L. Ruggiero and D.~Astesiano, ``Gravitational lensing observables in
  stationary and axisymmetric solutions in general relativity,'' {\em Phys.
  Rev. D}, vol.~112, p.~104044, Nov 2025.

\bibitem{Astesiano:2022gph}
D.~Astesiano, ``{Rigid rotation in GR and a generalization of the virial
  theorem for gravitomagnetism},'' {\em Gen. Rel. Grav.}, vol.~54, no.~7,
  p.~63, 2022.

\end{thebibliography}

\end{document}